\documentclass[aps,preprint]{revtex4}%
\usepackage{amsfonts}
\usepackage{amsmath}
\usepackage{amssymb}
\usepackage{graphicx}
\usepackage[usenames]{color}%
\providecommand{\U}[1]{\protect\rule{.1in}{.1in}}

\begin{document}
\preprint{ }
\title[Vortices in a rotating Fermi gas]{Finite temperature vortices in a rotating Fermi gas}
\author{S. N. Klimin}
\altaffiliation{Department of Theoretical Physics, State University of Moldova}

\affiliation{TQC, Universiteit Antwerpen, Universiteitsplein 1, B-2610 Antwerpen, Belgium}
\author{J. Tempere}
\altaffiliation{Lyman Laboratory of Physics, Harvard University}

\affiliation{TQC, Universiteit Antwerpen, Universiteitsplein 1, B-2610 Antwerpen, Belgium}
\author{N. Verhelst}
\affiliation{TQC, Universiteit Antwerpen, Universiteitsplein 1, B-2610 Antwerpen, Belgium}
\author{M. V. Milo\u{s}evi\'{c}}
\affiliation{Departement Fysica, Universiteit Antwerpen, Groenenborgerlaan 171, B-2020
Antwerpen, Belgium}

\begin{abstract}
Vortices and vortex arrays have been used as a hallmark of superfluidity in
rotated, ultracold Fermi gases. These superfluids can be described in terms of
an effective field theory for a macroscopic wave function representing the
field of condensed pairs, analogous to the Ginzburg-Landau theory for
superconductors. Here, we have established how rotation modifies this
effective field theory, by rederiving it starting from the action of Fermi gas
in the rotating frame of reference. The rotation leads to the appearance of an
effective vector potential, and the coupling strength of this vector potential
to the macroscopic wave function depends on the interaction strength between
the fermions, due to a renormalization of the pair effective mass in the
effective field theory. The mass renormalization derived here is in agreement
with results of functional renormalization group theory. In the extreme BEC
regime, the pair effective mass tends to twice the fermion mass, in agreement
with the physical picture of a weakly interacting Bose gas of molecular pairs.
Then, we use our macroscopic wave function description to study vortices and
the critical rotation frequencies to form them. Equilibrium vortex state
diagrams are derived, and they are in good agreement with available results of
the Bogoliubov -- De Gennes theory and with experimental data.

\end{abstract}
\date{\today}
\maketitle

\section{Introduction \label{sec:intro}}

Vortices and vortex matter in superconductors and superfluid atomic gases have
been subjects of a great interest since a long time \cite{Bloch2008}. Stable
vortices in superconductors appear under the presence of an external magnetic
field. In superfluid atomic Bose and Fermi gases, vortices are stabilized when
a trapped gas rotates, because the superfluid cannot support rigid-body
rotation \cite{Stoof,Iskin2009}. Stable vortices and vortex arrays were
successfully generated experimentally in condensates of bosonic
\cite{Matthews1999,Madison2000,Raman2001,Abo2001} and fermionic cold atoms
\cite{Zwierlein2005}.

The experimental progress stimulated theoretical efforts to describe physics
of vortex formation in rotating trapped quantum gases. Different theoretical
methods were applied to describe the physics of the vortex matter in these
systems: the Gross-Pitaevskii (GP) equation for Bose gases
\cite{Fetter2009,Tsubota2002}, the Ginzburg - Landau (GL) formalism
\cite{Bruun}, the Bogoliubov - De Gennes (BdG) theory
\cite{Machida,Sensarma2006,Chien,Simonucci2013} and the superfluid density
functional theory \cite{Bulgac2013}\ for Fermi gases. The first calculation of
the critical rotational velocity for a trapped Fermi gas has been performed in
Ref. \cite{Bruun} using a thermodynamic calculation of the energy of a vortex
state. A similar calculation for the Bose gases was performed earlier in Ref.
\cite{Dalfovo}. The rotating Fermi condensates were investigated using
macroscopic hydrodynamic equations in Refs. \cite{Cozzini,Urban}. In Refs.
\cite{Warringa,Warringa2,BdG-Wei2012}, the vortex formation in a rotating
trapped Fermi gas is studied using the Bogoliubov - De Gennes (BdG) equations.
In Ref. \cite{Simonucci2015}, vortex arrays in rotating Fermi condensates are
analyzed using the coarse graining method for the BdG equations developed in
Ref. \cite{Simonucci2014}, and referred to as a local phase density
approximation (LPDA) to the BdG equations.

The BdG theory has been successfully extended to superfluid Fermi gases in the
whole BCS-BEC crossover. However, from the computational point of view, the
solution of the BdG equations for the fermionic wave functions is far more
involved than the solution of, e. g., the Gross-Pitaevskii equation or any
similar effective field approach describing the superfluid through a
macroscopic wave function. As a result, the application of the BdG formalism
is mostly limited to the zero-temperature properties of single-vortex states
\cite{Machida,Sensarma2006,Chien}. To circumvent this limitation there has
been a great interest in the development of effective field theories (EFT)
which describe a superfluid system in terms of a macroscopic wave function
(order parameter). The common key approximation for all branches of the EFT is
the gradient expansion of the pair field, assuming it to be slowly varying in
time and space. For example, the Ginzburg-Landau (GL) and the Gross-Pitaevskii
(GP) theories can be considered as versions of the EFT, which are applicable
in different ranges of parameters.

Effective field theories have been established for different cases in a number
of works, see, e. g., Refs.
\cite{Nishida,Marini1998,Schakel,KTD2014,Simonucci2014}, and used to describe
non-uniform excitations (e. g., vortices, solitons) in Fermi gases in the
BCS-BEC crossover. A notable example is the coarse-grained approximation to
the static BdG formalism of Ref. \cite{Simonucci2014}, which allowed to extend
the analysis to the whole temperature range below $T_{c}$.

The present study is based on the finite-temperature EFT for quantum gases in
the BCS-BEC crossover formulated in our previous works. \cite{KTD,KTLD2015}.
This development of the EFT, based on a gradient expansion of the pairing
order parameter at finite temperatures, is dynamic, accounting for both
first-order and second-order time derivatives of the pair field. This allows
to treat both equilibrium and time-dependent phenomena in superfluid Fermi
gases. The gradient expansion is a common intrinsic element of an EFT.
Therefore all advantages and shortcomings of this approach are not specific to
the present work but are common for all EFTs (including GL and GP). Our
derivation of the basic expressions of the finite temperature EFT
\cite{KTD,KTLD2015} is based on a straightforward extension of the
first-nonvanishing-order expansion of the pair field action in powers of the
pair field $\Psi$ by a complete exact summation of the series in powers of
$\Psi$. It does not contain any additional hypothesis or model with respect to
the well-established EFT derived previously for quantum gases in the BCS-BEC
crossover near $T_{c}$, e. g., in Refs.
\cite{Nishida,Marini1998,Schakel,deMelo1993,Diener2008}. The
finite-temperature EFT has been tested by several successful applications to
quantum gases \cite{KTLD2015,KTD2014,LAKT2015} which confirmed its validity.

The method described in Refs. \cite{KTD,KTLD2015} was applied to solitons in a
fermionic superfluid \cite{KTD2014}, where its advantage becomes clear: an
analytic solution to the field equation is available. This approach will be
referred to as KTD, and shown to compare successfully to the BdG formalism in
the appropriate limit. Comparing this result to the numerical BdG simulations
has shown that the effective field theory of \cite{KTD,KTLD2015} is applicable
throughout the BCS-BEC crossover except for the combination of the BCS regime
and temperatures far below $T_{c}$ \cite{LAKT2015}, as expected (see the
corresponding discussion in Ref. \cite{Simonucci2014}).

In order to clearly indicate the place of the present work in the scientific
context, we stress that extensions the BdG and Gor'kov theories which embrace
BCS to BEC regimes for cold quantum gases were developed before in many works
starting from the Nozi\`{e}res and Schmitt-Rink (NSR) scheme, see, e. g.,
Refs. \cite{NSR,deMelo1993}. Within the Gaussian pair fluctuation
approximation (GPF), the path integral description of BCS to BEC crossover
treats the pairing channel at the level of the saddle point, and the Gaussian
fluctuations are incorporated into a renormalized chemical potential. There is
no real feedback of these fluctuations to the saddle point results as noted in
Ref. \cite{Diener2008}. However the EFT that we developed in Refs.
\cite{KTD,KTLD2015} is not completely equivalent to the GPF approach. We go
beyond GPF in what concerns the amplitude of the fluctuations: it is not
assumed to be small.

A complementary approach for Fermi gases in the BCS-BEC crossover is based on
the BCS-Leggett ground state \cite{Leggett,Perali}. The main difference
between these two methods is that the NSR-based scheme reaches the BCS-BEC
crossover by starting from the BEC limit, and the BCS-Leggett based scheme
reaches this crossover by starting from the BCS limit (for a detailed
comparison, see Ref. \cite{Levin}). Our recent works
\cite{KTD,KTLD2015,KTD2014,LAKT2015} lie within the context of the former.

The new elements of our version of the EFT, and, particularly, the message of
the present paper can be described as follows. The GL approach with
microscopically derived coefficients uses the pair field as a small parameter.
Therefore it is valid only near $T_{c}$. On one hand, the extension of the GL
approach for quantum gases valid near $T_{c}$ in the whole BCS-BEC crossover
and at $T=0$ in the BEC limit was developed in Ref. \cite{deMelo1993}. On the
other hand, an all-temperature extension of the GL method for BCS
superconductors was developed by Tewordt and Werthammer
\cite{Tewordt,Werthammer} using the gradient expansion for the order
parameter. Our recent treatment \cite{KTD,KTLD2015} partly fills an existing
gap, finding a similar extension of the GL method for quantum gases in the
BCS-BEC crossover.

Finally, the specific message of the present paper is an incorporation of
rotation into the effective field theory of Refs \cite{KTD,KTLD2015}. This is
done in Sec. \ref{sec:EFT} by including the rotating potential at the level of
the fermionic degrees of freedom, and deriving the modified EFT for the
macroscopic wave function. As we will show, the vector potential of rotation
contains the renormalization factor for the pair mass, which is in agreement
with results of the functional renormalization group theory \cite{Diehl}. In
Sec. \ref{sec:Results}, we show the equilibrium vortex state diagrams and
determine the critical rotation frequencies as a function of temperature and
interaction strength, compare the results with those of Refs.
\cite{Warringa,Warringa2,Simonucci2015} and analyze their connection with the
experimental data \cite{Zwierlein2005}. Our results are summarized in Sec.
\ref{sec:Conclusions}.

\section{Effective field action \label{sec:EFT}}

In the present work, we consider a rotating Fermi gas confined to an
anisotropic parabolic trap described by the confinement frequencies
$\omega_{j}$ ($j=x,y,z$) within the KTD approach described in Refs.
\cite{KTD,KTLD2015} and based on the path-integral description of the
interacting Fermi gas. The Hubbard-Stratonovich transformation is used to
introduce the bosonic pair field $\Psi$, and the action functional for these
fields is obtained by integrating out the fermionic degrees of freedom. In the
resulting action, a gradient expansion is performed, not around $\Psi=0$ as in
the Ginzburg - Landau approach, but around the coordinate-dependent
saddle-point value to be determined self consistently. The bosonic pair field
is then interpreted as a macroscopic wave function for the superfluid pair condensate.

The regimes of validity of this method have been studied in detail in Ref.
\cite{LAKT2015}. It is relevant to discuss once more the criteria of validity
of the EFT in the present work. A necessary condition for the validity of this
approach is the same as that for known effective field methods, e.~g., the
Ginzburg-Landau and Gross-Pitaevskii formalisms: the bosonic field $\Psi$ must
vary sufficiently slowly in space and in time. This condition is consistent
with a large number of particles in the superfluid system. Therefore we
restrict the treatment to Fermi gases with a sufficiently large number of
particles or sufficiently strong coupling in order to ensure $R_{j}\gg\xi$ and
$R_{c,j}\gg\xi$, where $R_{j}=\left(  \hbar/\left(  m\omega_{j}\right)
\right)  ^{1/2}$ is the characteristic scale for the trap potential along the
$j$-th axis, $R_{c,j}$ is the size of the superfluid cloud along the same
axis, and $\xi$ is the characteristic scale of non-uniform excitations. The
parameter $\xi$ can then be interpreted as the healing length for these
excitations, e. g., vortices or solitons.

In order to determine the range of applicability of the EFT, other length
scales must also be taken into account, such as the particle spacing, the
scattering length, and the pair size. Two of them are crucial for the
criterion of applicability for effective field approaches: the healing length
and the pair size. The latter one can be estimated through the pair coherence
length $\xi_{pair}$ which was determined in
\cite{THPistolesiStrinati,THPalestiniStrinati} through the pair correlation
function of the fermion field operators $\psi_{\sigma}\left(  \mathbf{r}%
\right)  ,\psi_{\sigma}^{\dag}\left(  \mathbf{r}\right)  $,%
\begin{align}
g_{\uparrow\downarrow}(\mathbf{r})  &  =-\left(  \frac{n}{2}\right)  ^{2}+\\
&  +\left\langle \psi_{\uparrow}^{\dag}\left(  \mathbf{R}+\frac{\mathbf{r}}%
{2}\right)  \psi_{\downarrow}^{\dag}\left(  \mathbf{R}-\frac{\mathbf{r}}%
{2}\right)  \psi_{\downarrow}\left(  \mathbf{R}-\frac{\mathbf{r}}{2}\right)
\psi_{\uparrow}\left(  \mathbf{R}+\frac{\mathbf{r}}{2}\right)  \right\rangle
,\nonumber
\end{align}
using the definition
\begin{equation}
\xi_{pair}=\sqrt{\frac{\int d\mathbf{r}~r^{2}g_{\uparrow\downarrow}%
(\mathbf{r})}{\int d\mathbf{r}~g_{\uparrow\downarrow}(\mathbf{r})}}.
\end{equation}

Effective field approaches are applicable when the pair size is small with
respect to the size of a non-uniform solution itself, i. e., when $\xi
_{pair}\ll\xi$ (see also the similar discussion in Ref. \cite{Simonucci2014}).
As found in Ref. \cite{LAKT2015}, the domain of applicability of the KTD
effective field theory is extended with respect to the GL approach (valid at
$T$ close to $T_{c}$) towards low temperatures, and with respect to the GP
approach (valid in the BEC limit) towards BCS. The KTD effective field theory
is thus not valid in the BCS regime combined with low temperatures $T\ll
T_{c}$.

In order to incorporate rotation into the KTD approach, we first consider the
single-particle Hamiltonian for a fermionic atom with mass $m$ confined to an
anisotropic parabolic trap in the rotating frame of reference. The rotation
leads to the appearance of the term $\left(  -\omega\hat{L}_{z}\right)  $
where $\omega$ is the rotation frequency and $\hat{L}_{z}$ is the $z$
component of the orbital angular momentum of the particle. Therefore the
single-particle Hamiltonian in the rotating frame of reference is
\cite{Gao2006,Urban2008}:%
\begin{equation}
H=-\frac{\left(  \nabla-i\mathbf{A}\left(  \mathbf{r}\right)  \right)  ^{2}%
}{2m}+\frac{m\left(  \omega_{\perp}^{2}-\omega^{2}\right)  }{2}\left(
x^{2}+y^{2}\right)  +\frac{m\omega_{z}^{2}}{2}z^{2}, \label{H4}%
\end{equation}
with the rotational\ vector potential \emph{for fermions},%
\begin{equation}
\mathbf{A}\left(  \mathbf{r}\right)  =m\left[  \boldsymbol{\omega}%
\times\mathbf{r}\right]  , \label{vecpot}%
\end{equation}
and the rotation vector
\begin{equation}
\boldsymbol{\omega}\equiv\omega\mathbf{e}_{z}.
\end{equation}
The effect of rotation in this Hamiltonian is explicitly subdivided to the
Coriolis and centrifugal contributions. The Coriolis contribution results in
the appearance of the vector potential (\ref{vecpot}), for which
$\nabla_{\mathbf{r}}\cdot A=0$. The centrifugal potential leads to the
softening of the confinement potential through $\omega_{\perp}^{2}%
\rightarrow\omega_{\perp}^{2}-\omega^{2}$. The trapped atomic configuration
can be stable when $\omega_{\perp}^{2}-\omega^{2}>0$. In the context of our
earlier assumption of a slowly varying field, the local density approximation
is suitable to take into account the confinement for a rotating Fermi gas
through a coordinate dependent chemical potential:%
\begin{equation}
\mu_{\omega}\left(  \mathbf{r}\right)  =\mu_{0}-\frac{m\left(  \omega_{\perp
}^{2}-\omega^{2}\right)  }{2}\left(  x^{2}+y^{2}\right)  -\frac{m\omega
_{z}^{2}}{2}z^{2}. \label{mu0}%
\end{equation}
This chemical potential enters the coordinate-dependent fermion density, which
is determined from the local number equation. Note that a parabolic
confinement potential facilitates the applicability of the effective field
theory and of the local density approximation with respect to a confinement
with sharp edges, e.~g., a box potential. Moreover, faster rotation makes the
confinement potential smoother, so that the rotation does not break up the
applicability of the present method. The local density approximation for
centrifugal and Coriolis contributions has, in general, the same range of
applicability as described above.

Within the present treatment, both the superfluid and normal components of the
Fermi gas are assumed to be in equilibrium in the rotating frame of reference.
This approximation is used in many works, see, e.~g., Refs. \cite{Warringa,
Warringa2,Simonucci2015} and references therein. Recently, it was argued that
rotation may cause a phase separation between a nonrotating superfluid core
and a rigidly rotating normal gas \cite{Iskin2009}. Also the cylindric
rotation symmetry about the $z$ axis is broken in experiments due to a
stirring field, which provides the rotation. The study of these effects is
however beyond the scope of the present work.

Within the path-integral formalism of preceding works
\cite{deMelo1993,Diener2008} and following to the scheme developed in Refs.
\cite{KTD,KTLD2015}, we start the treatment from the partition function of a
fermionic system determined by the path integral over the fermionic fields,
\begin{equation}
\mathcal{Z}\propto\int\mathcal{D}\left[  \bar{\psi},\psi\right]  e^{-S}.
\end{equation}
where the action functional $S$ is given by:%
\begin{equation}
S=\int_{0}^{\beta}d\tau\int d\mathbf{r}\left[  \sum_{\sigma=\uparrow
,\downarrow}\bar{\psi}_{\sigma}\left(  \frac{\partial}{\partial\tau}%
+H-\mu_{\sigma}\left(  \mathbf{r}\right)  \right)  \psi_{\sigma}+g\bar{\psi
}_{\uparrow}\bar{\psi}_{\downarrow}\psi_{\downarrow}\psi_{\uparrow}\right]  ,
\label{S}%
\end{equation}
where $\beta=1/\left(  k_{B}T\right)  $, $T$ is the temperature, and $k_{B}$
is the Boltzmann constant. To allow for spin imbalance in the Fermi gas,
chemical potentials $\mu_{\sigma}$ are introduced which can be different for
\textquotedblleft spin-up\textquotedblright\ and \textquotedblleft
spin-down\textquotedblright\ species. The coordinate dependent chemical
potentials $\mu_{\sigma}$ are determined by (\ref{mu0}) with $\mu
_{0}\rightarrow\mu_{0,\sigma}$ for each component. The interaction energy with
the coupling constant $g<0$ describes the model contact interactions between
fermions as, for example, in Ref. \cite{deMelo1993}. It represents the Cooper
pairing channel determined by the $s$-wave scattering between two fermions
with antiparallel spins. The one-particle Hamiltonian $H$ in the rotating
frame of reference is determined by formula (\ref{H4}).

A more detailed description of the derivation is given in Appendix A. After
the Hubbard-Stratonovich transformation which introduces the bosonic pair
fields $\left(  \bar{\Psi},\Psi\right)  $, integrating over the fermionic
fields, and the gradient expansion for the pair field with a complete
summation of the series in powers of $\left\vert \Psi\right\vert ^{2}$ in each
term of the gradient expansion, we arrive at the effective field action in the
rotating reference frame,%
\begin{align}
S_{eff}  &  =\int_{0}^{\beta}d\tau\int d\mathbf{r}\left\{  \left[  \Omega
_{s}\left(  w\right)  +\frac{D}{2}\left(  \bar{\Psi}\frac{\partial\Psi
}{\partial\tau}-\frac{\partial\bar{\Psi}}{\partial\tau}\Psi\right)  \right.
\right. \nonumber\\
&  +Q\frac{\partial\bar{\Psi}}{\partial\tau}\frac{\partial\Psi}{\partial\tau
}-\frac{R}{2w}\left(  \frac{\partial w}{\partial\tau}\right)  ^{2}+C\left(
\nabla_{\mathbf{r}}\bar{\Psi}\cdot\nabla_{\mathbf{r}}\Psi\right)  -E\left(
\nabla_{\mathbf{r}}w\right)  ^{2}\nonumber\\
&  \left.  \left.  +iD\mathbf{A}\cdot\left(  \bar{\Psi}\nabla_{\mathbf{r}}%
\Psi-\Psi\nabla_{\mathbf{r}}\bar{\Psi}\right)  \right]  \right\}  .
\label{SEFT}%
\end{align}
The coefficients of this effective field action and the thermodynamic
potential $\Omega_{s}$ are determined in Appendix A. They can depend on
coordinates through the squared amplitude of the pair field $w=\left\vert
\Psi\right\vert ^{2}$ and the chemical potentials $\mu=\left(  \mu_{\uparrow
}+\mu_{\downarrow}\right)  /2$ and $\zeta=\left(  \mu_{\uparrow}%
-\mu_{\downarrow}\right)  /2$. The linear term in the gradient expansion
appears due to rotation, because rotation breaks the local inversion symmetry.
Note that this linear term is derived in a straightforward way, without any ad
hoc assumption beyond the effective field approach.

It can be shown that the present approach is in agreement with
well-established results of the functional renormalization group theory
\cite{Boettcher,Diehl} in what concerns the effective pair mass. In the
microscopic theory of superconductivity \cite{Gorkov1958,Gorkov1959}, the pair
charge was determined as $e^{\ast}=2e$. As proven by Alben \cite{Alben1969},
the rotation of a superconductor brings a contribution to the vector potential
with the same charge/mass ratio for a pair as for a free electron. Therefore
the total vector potential of in the GL equation is twice the vector potential
for an electron, both for rotating and non-rotating superconductors. In
theories of rotating Fermi gases based on the GL or BdG equations
\cite{Gao2006,Simonucci2015}, this principle is kept. Contrary to the GL or
BdG based descriptions, effective field theories developed within the NSR-like
formalism \cite{Marini1998,Nishida,Schakel} and within the renormalization
group theory \cite{Boettcher,Diehl} necessarily contain the renormalized pair
effective mass $m_{p}^{\ast}$, which tends to $2m$ only in the extreme BEC
case. The present study lies within the latter of two aforesaid paradigms.
Hence we will arrive at a renormalized pair mass.

The derivation of the renormalized pair mass for Fermi gases in the BCS-BEC
crossover is described in Appendix A. It is shown that the renormalization
factor $\tilde{e}$ (associated with the ratio of the effective pair mass to
the fermion mass $\tilde{e}\equiv m_{p}^{\ast}/m$) is expressed through the
coefficients of the effective action (\ref{SEFT}) by:%
\begin{equation}
{\tilde{e}}=\frac{1}{C}\frac{\partial\left(  wD\right)  }{\partial w}.
\label{rf}%
\end{equation}

\begin{figure}
[th]
\begin{center}
\includegraphics[
height=4.5339in,
width=3.5081in
]%
{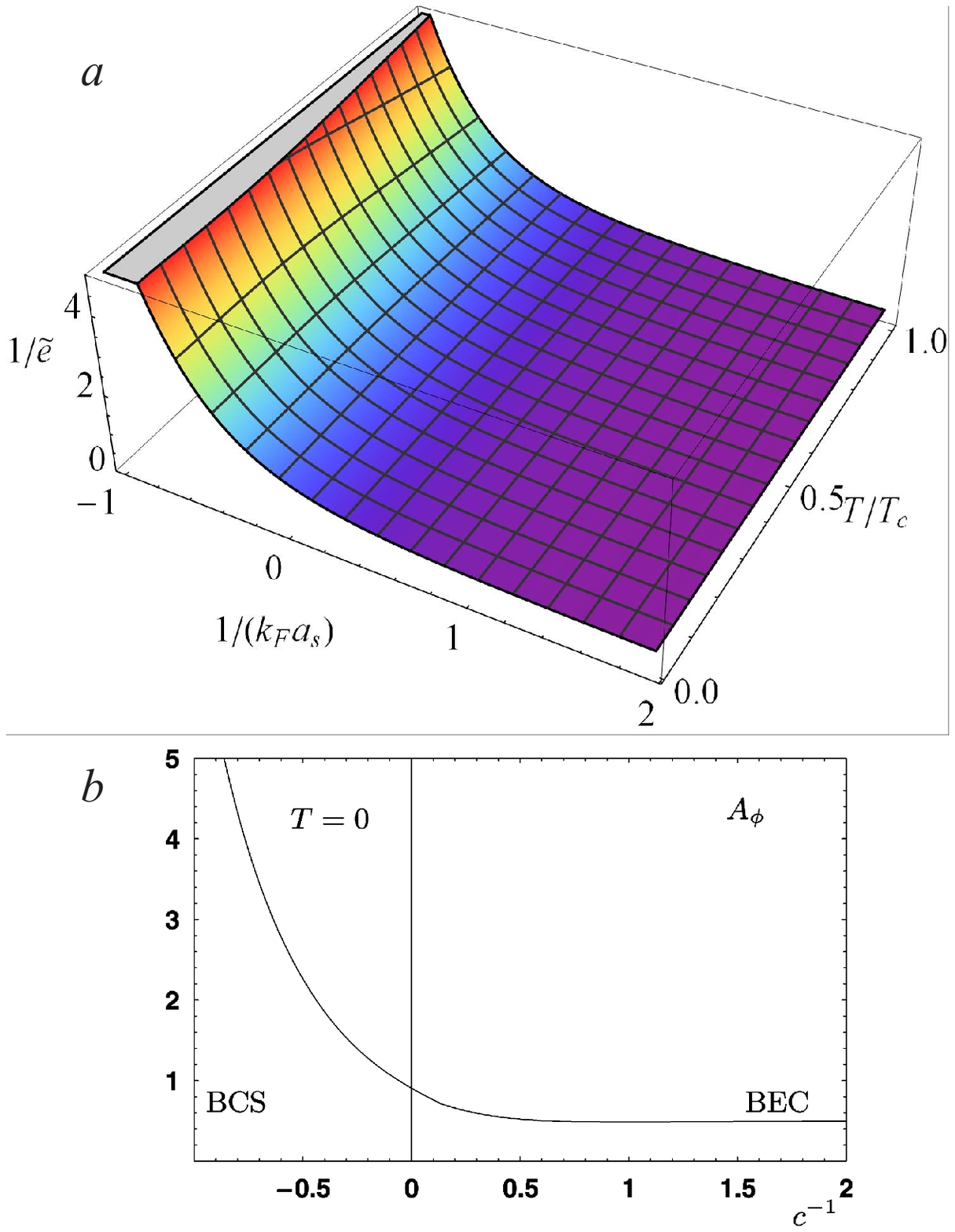}%
\caption{(Color online). (\emph{a}) Inverse renormalization factor\ $1/\tilde
{e}$ as a function of the dimensionless inverse scattering length $1/\left(
k_{F}a_{s}\right)  $ and the relative temperature $T/T_{c}$ for a
three-dimensional Fermi gas in a cylindricaly symmetric parabolic confinement
potential, with the number of particles per unit length $N=1000$. (\emph{b})
Inverse renormalization factor obtained within the functional renormalization
group theory \cite{Diehl} for $T=0$. The renormalization factor $\tilde{e}$ is
associated with the effective pair mass, as shown in Appendix A.}%
\label{fig:kappa}%
\end{center}
\end{figure}

Fig. \ref{fig:kappa} conveys the fact that the present EFT is in line with
well-established results of the functional renormalization group theory
\cite{Boettcher,Diehl}. Here, the inverse of the renormalization factor,
$1/\tilde{e}$, is plotted as a function of the inverse scattering length
$1/\left(  k_{F}a_{s}\right)  $ (where $k_{F}$ is the Fermi wave vector) and
the temperature, when $T$ passes from zero to $T_{c}$ for a three-dimensional
Fermi gas confined to a cylindrically symmetric parabolic confinement
potential, with the number of particles per unit length set to $N=1000$.%

As seen from Fig. \ref{fig:kappa}, the inverse renormalization factor only
slightly depends on the temperature, and tends to 1/2 in the BEC limit, where
the Fermi superfluid can be described as a Bose gas of molecules with the mass
$m_{p}^{\ast}=2m$. Moving away from the BEC\ limit, $1/\tilde{e}$ gradually
increases. The obtained behavior of the renormalization factor as a function
of the inverse scattering length is in good agreement with the prediction of
the functional renormalization group theory of Ref. \cite{Diehl}. This\ is one
of the key results of the present approach, which was not yet applied before
to rotating Fermi gases. Thus, besides a renormalization of the chemical
potential, an important element of the BCS-BEC crossover in the present work
is a renormalization of all coefficients of the effective field action,
including the renormalization of the pair mass.

Finally, the effective field action for a two-band system is straightforwardly
determined in the same way as in Ref. \cite{KTLD2015}. We obtain action
functionals for the separate fields, and a coupling given by an interband
Josephson term:%
\begin{equation}
S_{eff}^{\left(  2b\right)  }=\sum_{j=1,2}S_{eff}^{\left(  j\right)  }%
-\int_{0}^{\beta}d\tau\int d\mathbf{r~}\frac{\sqrt{m_{1}m_{2}}}{4\pi}%
\gamma\left(  \bar{\Psi}_{1}\Psi_{2}+\bar{\Psi}_{2}\Psi_{1}\right)  ,
\end{equation}
Here $S_{eff}^{\left(  j\right)  }$ is the single-band effective field action
for the $j$-th band determined by (\ref{SEFT}) with $j=1,2$, and $\gamma$ is
the strength of the interband coupling. As derived in Ref. \cite{KTLD2015},
the coupling parameter $\gamma$ is fixed by the interband scattering lengths,%
\begin{equation}
\gamma=2\left(  \frac{1}{a_{s,3}}-\frac{1}{a_{s,4}}\right)  ,
\end{equation}
where the scattering lengths $a_{s,3}$ and $a_{s,4}$ are related to the
interband scattering for the fermions with antiparallel and parallel spins, respectively.

\section{Vortex formation \label{sec:Results}}

In order to study the formation of vortices and vortex pairs in rotated
superfluid Fermi gases, we use the amplitude-phase representation for the pair
field similarly as in Refs. \cite{KTD2014,LAKT2015},
\begin{equation}
\Psi\left(  \mathbf{r}\right)  =\left\vert \Psi_{\infty}\right\vert a\left(
\mathbf{r}\right)  e^{i\theta\left(  \mathbf{r}\right)  }. \label{amfas}%
\end{equation}
In this expression, $\left\vert \Psi_{\infty}\right\vert $ is the uniform
background amplitude determined by solving gap and number equations for the
uniform system. The amplitude modulation (the \textquotedblleft
hole\textquotedblright\ in the modulus of the order parameter at the vortex
core) is modeled by the real function $a(\mathbf{r})$. The phase pattern is
taken into account by $\theta(\mathbf{r})$ -- for a vortex aligned with the
$z$-axis, this is the angle around the $z$-axis. With this representation for
$\Psi$, the free energy corresponding to the effective action becomes%
\begin{align}
F  &  =\int d\mathbf{r}\left\{  \left[  \Omega_{s}\left(  w\right)  +\frac
{1}{2}\rho^{\left(  qp\right)  }\left(  \nabla_{\mathbf{r}}a\right)
^{2}\right]  \right. \nonumber\\
&  \left.  +\frac{1}{2}\rho^{\left(  sf\right)  }a^{2}\left(  \nabla
_{\mathbf{r}}\theta-\mathring{e}\mathbf{A}\right)  ^{2}-\frac{1}{2}%
\rho^{\left(  sf\right)  }\left(  a\mathring{e}\mathbf{A}\right)
^{2}\right\}  , \label{F}%
\end{align}
with
\begin{align}
\rho^{\left(  sf\right)  }  &  =2C\left\vert \Psi_{\infty}\right\vert ^{2},\\
\rho^{\left(  qp\right)  }  &  =2\left(  C-4E\right)  \left\vert \Psi_{\infty
}\right\vert ^{2}.
\end{align}
The parameters $\rho^{\left(  sf\right)  }$ and $\rho^{\left(  qp\right)  }$
represent, respectively, the superfluid density and the quantum pressure
coefficient, as established in Refs. \cite{KTD2014,KTLD2015}. In order to find
the conditions of stability for the vortex solutions, we consider the
difference between two free energies:%
\begin{equation}
\delta F\equiv F_{vortex}-F_{0} \label{dF}%
\end{equation}
where $F_{vortex}$ and $F_{0}$ are given by (\ref{F}), respectively, with and
without vortices. The bounds for the equilibrium vortex state diagrams with
several vortex configurations are determined from the comparison of the free
energies corresponding to these configurations.

From here on, we focus on vortex stability conditions for a one-band Fermi gas
in three dimensions, trapped in a cylindrically symmetric parabolic potential
characterized by the confinement frequency $\omega_{0}$, and rotating around
the symmetry axis at a frequency $\omega$. We do not consider at the present
stage the case when the population imbalance $\zeta$ is other than zero. The
area of existence of vortices lies, in general, inside the area of existence
for a superfluid state in a rotating Fermi gas. The latter one extends from
the zero rotation frequency $\omega=0$ to a critical rotation frequency for
the superfluid state $\omega_{\max}<\omega_{0}$. For $\omega>\omega_{\max}$,
the system turns into the normal state
\cite{Urban2008,Bausmerth2008,Veillette2006}.%

\begin{figure}
[h]
\begin{center}
\includegraphics[
height=2.9611in,
width=3.7351in
]%
{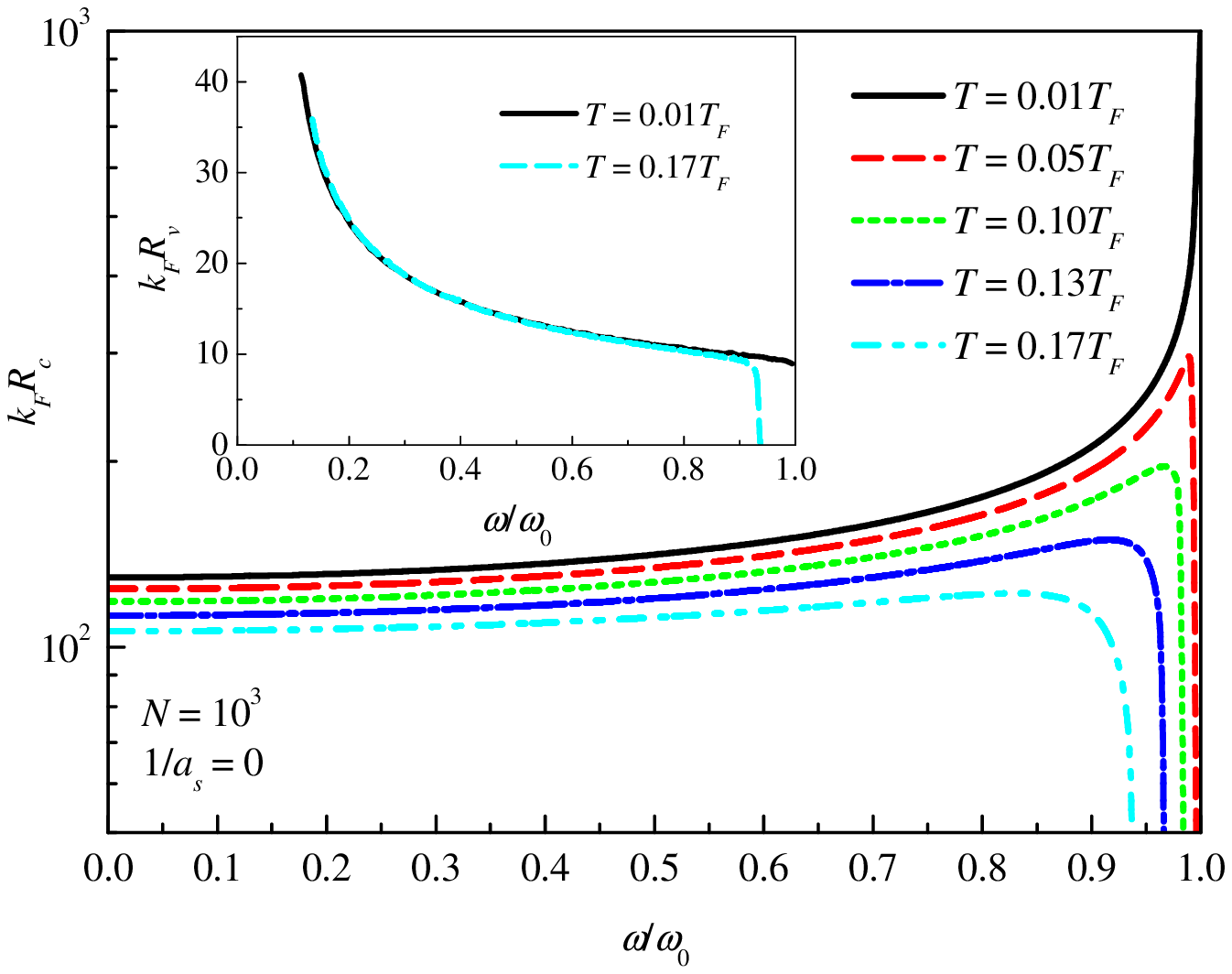}%
\caption{(Color online). Radius of the superfluid state as a function of the
rotation frequency for a rotating Fermi gas with $1/a_{s}=0$ and $N=10^{3}$
confined to a cylindrically symmetric parabolic potential, at different
temperatures. Inset shows the half-distance between vortex centers for a
vortex pair at two temperatures.}%
\label{fig:radii}%
\end{center}
\end{figure}

Fig. \ref{fig:radii} shows the behavior of the radius of the superfluid state
$R_{c}$ and the half-distance between vortex centers $R_{v}$ for a vortex pair
(inset) as a function of the relative rotation frequency $\omega/\omega_{0}$
for a rotating Fermi gas with $1/a_{s}=0$ and $N=10^{3}$ confined to a
cylindrically symmetric parabolic potential. The dependence of the radius of
the superfluid state versus $\omega$ is non-monotonic. When rotation gradually
becomes faster but $\omega$ is not yet very close to its critical value
$\omega_{\max}<\omega_{0}$ (where the superfluid state disappears), $R_{c}$
slowly increases, because the confinement weakens due to the centrifugal
force. When $\omega$ is sufficiently close to $\omega_{\max}$, the superfluid
core shrinks, turning to zero at $\omega=\omega_{\max}$. The critical value
$\omega_{\max}$ decreases with increasing temperature, in accordance with the
predictions of other works \cite{Warringa2,Veillette2006}.

Fig. \ref{fig:radii} allows us to see also the temperature dependence of the
size of the superfluid state and of the half-distance for the vortex pair.
When $\omega$ is not close enough to $\omega_{\max}$, the radius $R_{c}$
decreases rather slowly with rising temperature. In the vicinity of
$\omega_{\max}$, this decrease becomes much faster. The half-distance between
vortices for a pair weakly depends on the temperature, except near
$\omega_{\max}$, where $R_{v}$ falls together with $R_{c}$.

For a non-rotating Fermi gas and at sufficiently low rotation frequencies,
vortices are not stable as long as the free energy (\ref{F}) without vortices
is lower than the free energy with vortices. When increasing $\omega$,
vortices can become stable starting from a certain critical rotation frequency
$\omega=\omega_{c,1}$. There may exist also an upper critical rotation
frequency $\omega_{c,2}<\omega_{\max}$ such that the vortex state turns back
to the superfluid state for $\omega_{c,2}<\omega<\omega_{\max}$. The
appearance of an upper critical rotation frequency was also predicted by the
BdG theory \cite{Warringa2}. The existence of a superfluid without any vortex
at a fast rotation may seem counter-intuitive, but it has a transparent
physical explanation. As seen from Fig. \ref{fig:radii}, starting from
sufficiently large rotation frequencies, the radius of the superfluid state
decreases. When the size of the superfluid is of the same order as the vortex
size (or smaller), the formation of vortices can be not energetically
favorable. This explains the existence of a superfluid without vortices at a
fast rotation.%

\begin{figure}
[th]
\begin{center}
\includegraphics[
height=3.1622in,
width=3.7904in
]%
{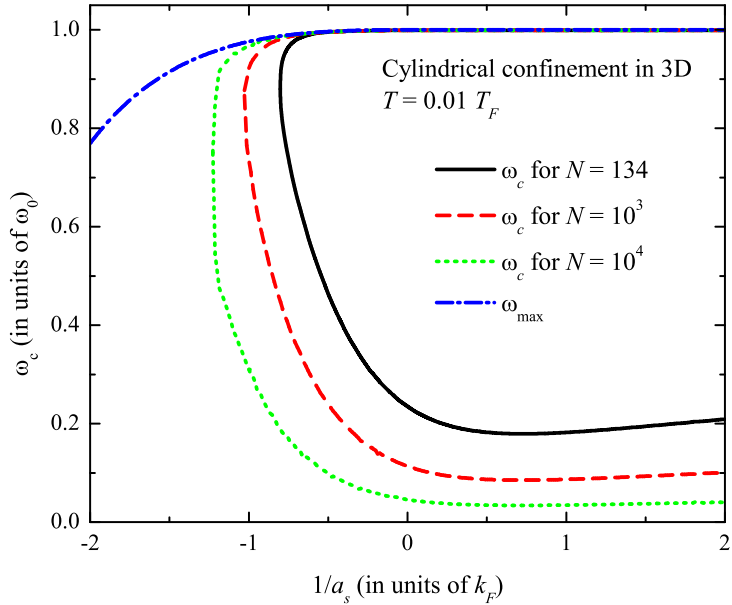}%
\caption{(Color online). Area of existence for vortices for a Fermi gas
trapped in a cylindrically symmetric parabolic potential at $T=0.01T_{F}$,
with different numbers of particles per unit length.}%
\label{fig:areav1}%
\end{center}
\end{figure}

The area of existence for vortices for a system with different numbers of
particles $N$ per unit length at $T=0.01T_{F}$ (where $T_{F}=E_{F}/k_{B}$) is
shown in Fig. \ref{fig:areav1}. When comparing our results with those of Ref.
\cite{Warringa2}, one should note different units for the number of particles
per unit length in that work than in the present treatment. Here, the lengths
are measured in units of $1/k_{F}$, and in Ref. \cite{Warringa2}, the unit
length is chosen as the oscillator length $l_{o}\equiv\left(  \hbar/\left(
m\omega_{0}\right)  \right)  ^{1/2}$, where $\omega_{0}$ is the confinement
frequency. We denote by $\mathcal{N}$ the number of particles per unit length
according to Ref. \cite{Warringa2}, and ours by $N$. Therefore these two
numbers are related to each other as $\mathcal{N}=N\mathcal{~}k_{F}l_{o}$. In
our units, $l_{o}=\left(  2/\omega_{0}\right)  ^{1/2}$ with $\omega_{0}%
=\sqrt{8/\left(  15\pi N\right)  }$, and hence $\mathcal{N}=\left(
15\pi/2\right)  ^{1/4}N^{5/4}$. In particular, the value $N=134$ corresponds
to $\mathcal{N}\approx1000$ in Ref. \cite{Warringa2}.

We do not perform a quantitative comparison of the {\normalsize equilibrium
vortex state diagrams} calculated within the current approach with those
obtained by the BdG method \cite{Warringa2} since the study in Ref.
\cite{Warringa2} has been performed for the BCS regime, while the quantitative
results of the current effective field theory, as discussed in Refs.
\cite{KTLD2015,KTD2014}, are hardly applicable in the BCS regime at $T\ll
T_{c}$. Nevertheless, the qualitative behavior of the boundary for the area of
stable vortices is in agreement with the predictions of the BdG theory even in
the BCS side. Particularly, we can see a bend-over of the critical rotation
frequency and hence the existence of both a lower and an upper critical
rotation frequency at weak coupling. At higher coupling strengths, the upper
critical rotation frequency for the vortex formation tends to the critical
rotation frequency for the superfluid state.

\begin{figure}
[th]
\begin{center}
\includegraphics[
height=2.8172in,
width=3.986in
]%
{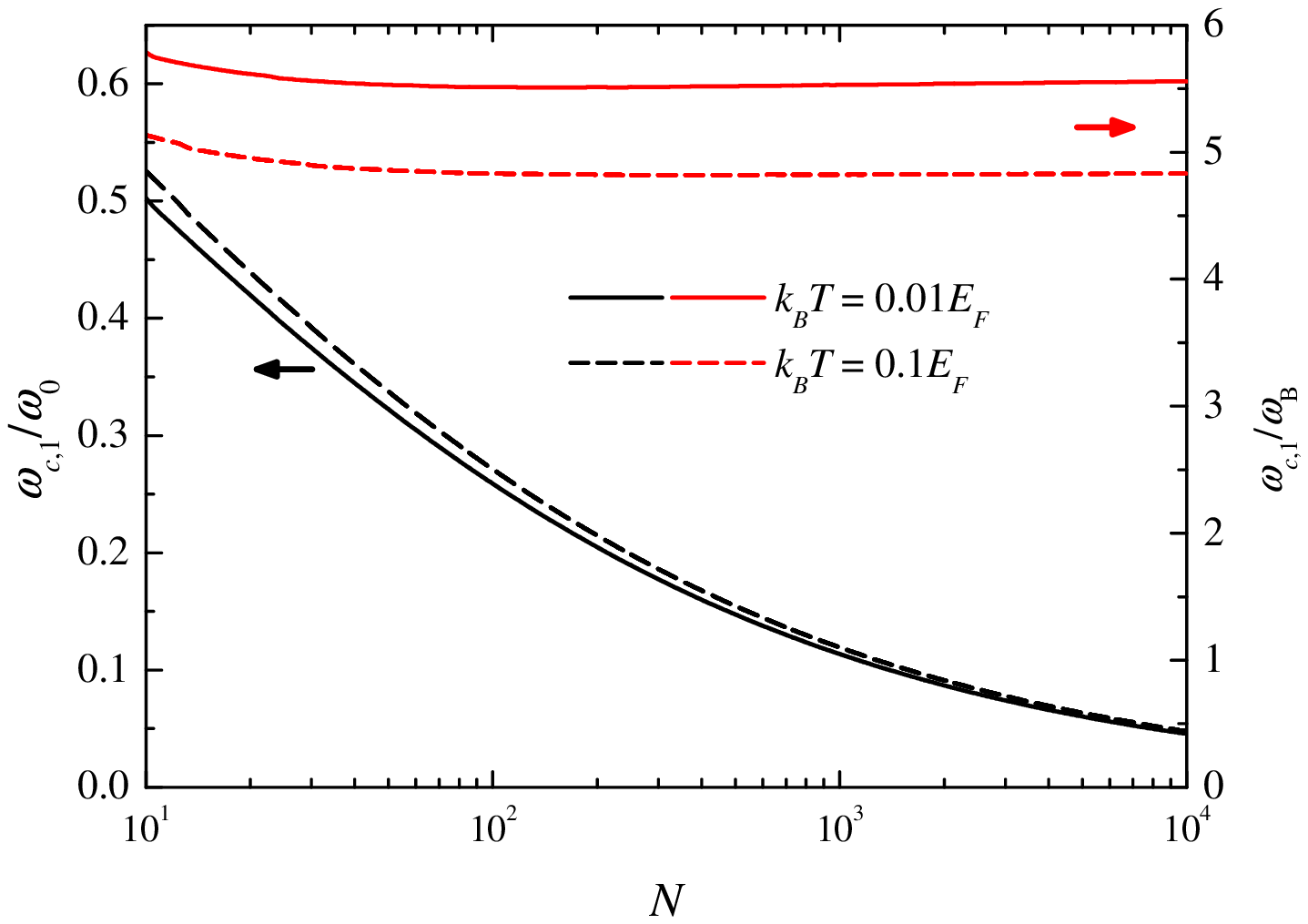}%
\caption{(Color online). Left-hand axis (indicated by the arrow for two lower
curves): the lower critical rotation frequency $\omega_{c,1}$ (in units of the
trapping potential parameter $\omega_{0}$) for a trapped Fermi gas as a
function of the number of particles per unit length for $1/\left(  k_{F}%
a_{s}\right)  =0$, at two temperatures $k_{B}T=0.01E_{F}$ and $k_{B}%
T=0.1E_{F}$. \emph{Right-hand axis} (indicated by the arrow for two upper
curves): the ratio $\omega_{c}/\omega_{B}$ where $\omega_{B}$ is given by
formula (\ref{trend}).}%
\label{fig:largeN}%
\end{center}
\end{figure}

The region of vortex stability extends deeper into the BCS side and to smaller
values of $\omega_{c,1}$ when increasing the number of particles. For
sufficiently large $N\gtrsim10^{4}$, stable vortices as predicted by the
current formalism can be observed in the entire experimentally available
BCS-BEC crossover region ($-1.2<1/\left(  k_{F}a_{s}\right)  <3.8$), in line
with the experimental observations \cite{Zwierlein2005}. We have checked
numerically that the lower critical rotation frequency $\omega_{c,1}$ for a
single vortex in a Fermi gas with a large number of particles behaves in
accordance with the estimation \cite{Bruun,Nygaard}:%
\begin{equation}
\left.  \omega_{c,1}\right\vert _{N\gg1}\propto\omega_{B}\equiv\frac{1}%
{R_{c}^{2}}\ln\left(  \frac{R_{c}}{\xi}\right)  , \label{trend}%
\end{equation}
where $R_{c}$ is the radius of the superfluid state in a trap, and $\xi$ is
the healing length which characterizes the vortex size. The result of this
numerical check is shown in Fig. \ref{fig:largeN}. It shows the lower critical
rotational frequency for a Fermi gas as a function of the number of particles
per unit length and the ratio of the critical frequency compared to the
analytic expression (\ref{trend}). We see that the ratio $\omega_{c,1}%
/\omega_{B}$ only slightly varies when $N$ passes from $N=10$ to $N=100$, so
that the asymptotic trend (\ref{trend}) is clearly visible already when $N$ is
not very large.%

A similar asymptotic dependence for a Fermi gas trapped to a 3D spherically
symmetric confinement potential was predicted in Ref. \cite{Bruun} for a Fermi
gas at zero temperature. In the present treatment, we find that the trend
(\ref{trend}) is kept also at finite temperatures.

We can compare the obtained critical rotation frequency with the LPDA results
of Ref. \cite{Simonucci2015}, using the parameters of the experimental setup
of Ref. \cite{Riedl2011} where the unitary Fermi gas [$1/\left(  k_{F}%
a_{s}\right)  =0$] is trapped to an elongated trap with the confinement
frequencies $\omega_{\perp}\approx2\pi\times680%
\operatorname{Hz}%
$ and $\omega_{z}\approx2\pi\times24%
\operatorname{Hz}%
$. When approximating this setup by a cylindrical confinement potential, we
arrive at the number of particles per unit length $N\sim10^{4}$. As seen from
Fig. \ref{fig:largeN}, for this number of particles, $\omega_{c,1}%
\approx0.045$, which is in good agreement with the lower critical rotation
frequency obtained in Ref. \cite{Simonucci2015}.

\begin{figure}
[h]
\begin{center}
\includegraphics[
height=3.5672in,
width=3.6188in
]%
{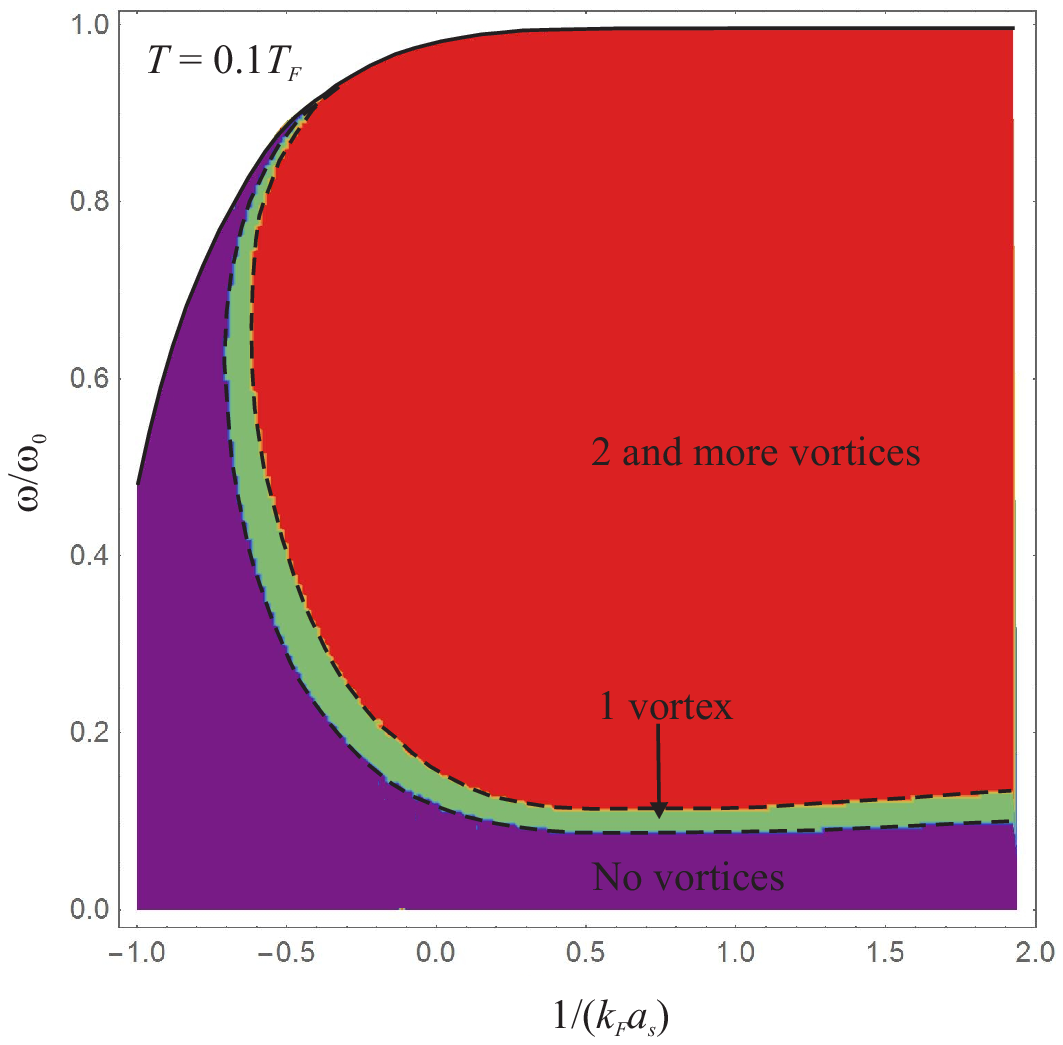}%
\caption{(Color online). Equilibrium vortex state diagram for a trapped
rotating Fermi gas in a cylindrically symmetric parabolic confinement
potential, showing the critical rotation frequencies as a function of the
inverse scattering length for $T=0.1T_{F}$ and the number of particles per
unit length $N=10^{3}$. The critical rotation frequencies are plotted for a
single vortex and for a vortex pair. Also the upper bound for the rotation
frequency is shown, which restricts the area of existence for the superfluid
state.}%
\label{fig:areav2}%
\end{center}
\end{figure}

When the rotation frequency is increased beyond $\omega_{c,1}$, a second
vortex may enter the superfluid. In the equilibrium vortex state diagram of
Fig. \ref{fig:areav2}, we distinguish the superfluid states with no vortex,
one vortex and two or more vortices, in a trapped Fermi gas with $N=1000$ at
the temperature $T=0.1T_{F}$. This temperature is higher than that for Fig.
\ref{fig:areav1}, and as a consequence the BCS-side boundary for vortex
formation is found to shift to stronger coupling strengths. The boundary
between the regimes with one and two vortices behaves similarly to the
critical rotation frequency for a single vortex. It also exhibits a bend-over.
The lower critical rotation frequency for a vortex pair is higher than the
lower critical rotation frequency for a single vortex. On the contrary, the
upper critical rotation frequency for a vortex pair is lower than the higher
critical rotation frequency for a single vortex. Also the weak-coupling bound
of $1/a_{s}$ for a single vortex lies more towards the BCS side with respect
to that for a vortex pair. Thus the area where two or more stable vortices can
exist lies entirely inside the area of stability for a single vortex.

\begin{figure}
[th]
\begin{center}
\includegraphics[
height=4.0321in,
width=5.985in
]%
{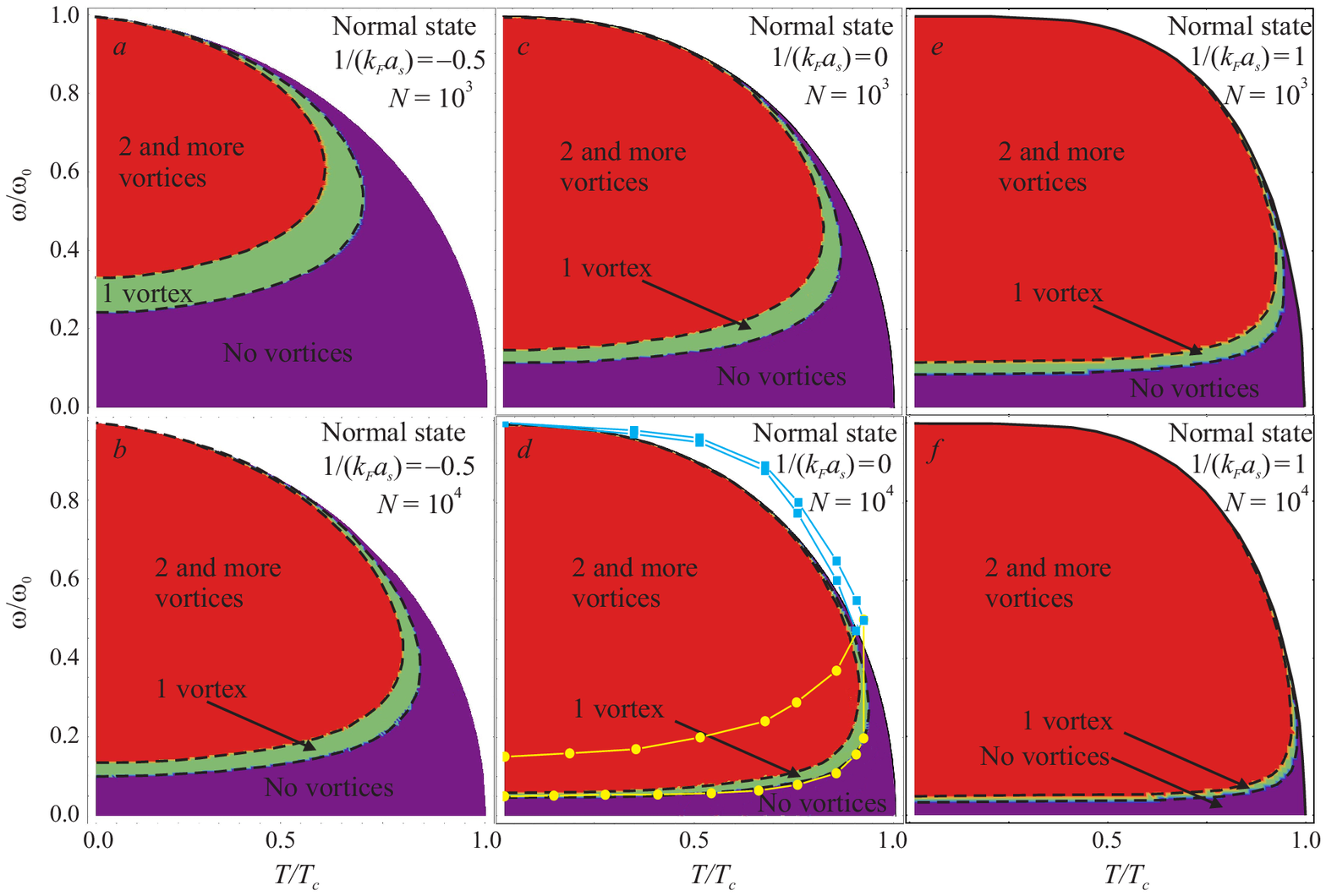}%
\caption{(Color online). Equilibrium vortex state diagrams for a trapped
rotating Fermi gas in a cylindrically symmetric parabolic confinement
potential, showing the critical rotation frequencies as a function of the
temperature for two numbers of particles per unit length and three inverse
scattering lengths (indicated in the figure). The notations are the same as in
Fig. \ref{fig:areav2}. Symbols in panel (\emph{d}) show the critical rotation
frequencies $\omega_{c,1}$ (dots) and $\omega_{c,2}$ (squares) from the
Supplement to Ref. \cite{Simonucci2015}.}%
\label{fig:areav3}%
\end{center}
\end{figure}

In Fig. \ref{fig:areav3}, we plot the equilibrium vortex state diagrams as a
function of the variables $\left(  \omega/\omega_{0},T/T_{c}\right)  $ for two
numbers of particles per unit length $N=10^{3}$ and $N=10^{4}$, and for three
values of the inverse scattering length $1/\left(  k_{F}a_{s}\right)  =-0.5$
(the BCS case), $1/\left(  k_{F}a_{s}\right)  =0$ (unitarity), and $1/\left(
k_{F}a_{s}\right)  =1$ (the BEC case). It should be noted that different
areas\ in the equilibrium vortex state diagrams do not refer to genuine
thermodynamical phases, which are superfluid and normal phases. Also,
equilibrium vortex state diagrams in a uniform superfluid (like $^{3}$He)
would be different from those in a trapped Fermi gas.%

In this equilibrium vortex state diagram, the transition lines between the
regimes with no vortex, one vortex, and two or more vortices bend over leading
to reentrant behavior of the critical rotation frequencies as a function of
temperature. This reentrant dependence has a clear physical sense. On one
hand, at higher temperatures, the radius of the superfluid phase (which is
surrounded by the normal phase) decreases. On the other hand, the healing
length, which determines the vortex size, increases when the temperature rises
towards $T_{c}$. When the healing length is sufficiently large, the existence
of stable vortices becomes energetically non-favorable with respect to the
superfluid state. The obtained equilibrium vortex state diagrams exhibit a
clear similarity to those obtained in Ref. \cite{Warringa2} (where they are
calculated in the far BCS regime and at lower temperatures than those
considered in the present work) and in Ref. \cite{Simonucci2015}. When moving
from the BCS to the BEC regime, and when increasing the number of particles,
the area for a single vortex, as well as the area for a superfluid state
without vortices, become gradually narrower.

In Fig. \ref{fig:areav3}, the temperature is measured in units of the critical
temperature $T_{c}$ at the zero rotation. The critical temperatures calculated
using the background chemical potential in the mean-field approach are
overestimated with respect to experimental data, e. g., the mean-field value
at unitarity $T_{c}\approx0.4T_{F}$, while in the experiment \cite{Horikoshi},
$T_{c}\approx0.17T_{F}$. Taking Gaussian fluctuations into account
\cite{Haussmann} results in $T_{c}\approx0.21T_{F}$ in better agreement with
experimental estimate \cite{Horikoshi} for the critical temperature. However,
this will not qualitatively change the equilibrium vortex state diagrams when
$T$ is scaled to $T_{c}$.%

The equilibrium vortex state diagram shown in Fig. \ref{fig:areav3}(d)
corresponds to the same experimental setup as in Ref. \cite{Riedl2011},
theoretically considered in Ref. \cite{Simonucci2015}. For comparison, we plot
there also the critical rotation frequencies $\omega_{c,1}$ and $\omega_{c,2}$
from Fig. S2 of the Supplement to Ref. \cite{Simonucci2015}, shown by symbols.
The calculations in the present work are performed for a cylindrical
confinement, which only approximately simulates an elongated trap considered
in Ref. \cite{Simonucci2015}. Thus we expect only a qualitative agreement
between our results and those of Ref. \cite{Simonucci2015}. However, the
critical rotation frequencies in Fig. \ref{fig:areav3}(\emph{d}) appear to be
close to those in the equilibrium vortex state diagram calculated within LPDA
\cite{Simonucci2015}. It is also worth to note a good agreement between the
KTD effective field theory and LPDA on the upper critical temperature for the
vortex formation, as seen from Fig. \ref{fig:areav3}(\emph{d}).

There are also some differences between the critical rotation frequencies
derived within these two approaches. In the BdG method, there are two
definitions of the lower critical rotation frequency. A lower value of
$\omega_{c,1}$ corresponds to the critical angular frequency at which an
isolated vortex placed initially close to the trap center is attracted toward
the trap center, while the upper value of $\omega_{c,1}$ corresponds to the
critical rotation frequency at which an isolated vortex placed initially at
the edge is attracted toward the trap center. This appearance of different
critical rotation frequencies is apparently related to the fact that the LPDA
equation determines a dynamic stability of vortices. In the EFT, the condition
for the vortex formation follows from the comparison of the free energies with
and without vortices. In other words, we consider only the thermodynamic
stability of the vortex configurations. Therefore a single critical rotation
frequency is obtained in the present work. The upper value of $\omega_{c,1}$
can thus correspond to a thermodynamically metastable configuration. As soon
as the experimental preparation of the states of quantum atomic gases and the
measurements are performed during a finite time, both thermodynamically stable
and metastable configurations can be observable. Which critical rotation rate
is relevant for a particular experiment depends on the way in which the
experiment is performed.

According to the results shown in Figs. \ref{fig:largeN} and \ref{fig:areav3}%
(\emph{d}), the critical rotation frequency $\omega_{c,1}$ given by the
present EFT is in excellent agreement with the lowest of two values of
$\omega_{c,1}$ provided by the coarse grained BdG theory \cite{Simonucci2015}.
This may shed light on which of the two values of $\omega_{c,1}$ indicated in
Ref. \cite{Simonucci2015} corresponds to the thermodynamically stable state:
the lower one is stable while the higher one can be thermodynamically metastable.

A comparison with the observations of vortices in the experiment of Ref.
\cite{Zwierlein2005} indicates that the ranges of applicability of the BdG
formalism combined with the Thomas-Fermi approximation
\cite{Warringa,Warringa2} and the KTD effective field theory are complementary
to each other. The KTD field theory becomes more accurate towards the BEC
regime \cite{KTLD2015}, while, as concluded in Ref. \cite{Warringa2}, the BdG
method is quantitatively more reliable towards the BCS regime. It was found in
Refs. \cite{Warringa,Warringa2} that within the BdG theory, vortices in
rotating Fermi gases are formed only for relatively large negative scattering
lengths. On the contrary, the current formalism predicts the formation of
stable vortices in rotating Fermi gases in the whole BCS-BEC crossover, in
agreement with the experimental observations \cite{Zwierlein2005}.

The inverse scattering length was varied in the experiment of Ref.
\cite{Zwierlein2005} in a wide range from $1/\left(  k_{F}a_{s}\right)  =-1.2$
to $1/\left(  k_{F}a_{s}\right)  =3.8$, and vortices were observed in the
whole range of $1/\left(  k_{F}a_{s}\right)  $ between those values. In the
experiment \cite{Zwierlein2005}, $^{6}$Li atoms were trapped in an
approximately parabolic trap with the confinement frequencies $\omega_{\perp
}\approx2\pi\times57%
\operatorname{Hz}%
$ and $\omega_{z}\approx2\pi\times23%
\operatorname{Hz}%
$. This gives us an estimation of the trap length along the $z$ axis
$l_{z}\equiv\left(  \hbar/\left(  m\omega_{z}\right)  \right)  ^{1/2}%
\approx8.5%
\operatorname{\mu m}%
$. The total number of atoms was $N\propto10^{6}$. Thus we can estimate the
number of particles per unit length in order to qualitatively match the
experiment as $N\propto N/l_{z}\propto10^{4}$. The highest number of vortices
at a given stirring frequency was obtained at $1/\left(  k_{F}a_{s}\right)
\approx0.35$ which is rather close to the position of the minimum of the
critical rotation frequency for $N=10^{4}$ in Fig. \ref{fig:areav1}. It is
hard to extract the critical rotation frequency for a single vortex from the
experimental data of Ref. \cite{Zwierlein2005}. However, it is suggestive that
the minimum of the critical rotation frequency and the maximum of vortices at
a given (higher) rotation frequency lie close each other. Thus the above
results of the present work are in line with the experiment
\cite{Zwierlein2005} in what concerns the most favorable scattering length for
the vortex formation in a rotating Fermi gas. Also, the estimate of the
optimal rotation frequency within the modified finite temperature EFT is in a
good agreement with the result of the coarse grained BdG theory
\cite{Simonucci2015} and with the experiment \cite{Zwierlein2005}. This
agreement is remarkable despite the fact that the rotation is incorporated in
the LPDA equation of Ref. \cite{Simonucci2015} and in the present work in
different ways.

\section{Conclusions \label{sec:Conclusions}}

In the present work, we extend the effective field theory developed in Refs.
\cite{KTD,KTLD2015} for fermionic superfluids to the case of rotating Fermi
gases. The treatment is performed within the same path integral formalism as
in the theoretical studies of cold quantum gases which embrace BCS to BEC
regimes, performed in preceding works. The new physics in our recent works on
the EFT is related to an extension of the GL theory below $T_{c}$ in the
BCS-BEC crossover.

The rotation has been incorporated in the effective field action in a
straightforward way, leading to the appearance of an effective vector
potential as in other effective field theories. Therefore the physical
picture, e. g., for the formation of vortices, is qualitatively one and the
same in different formalisms (see, e. g., \cite{Warringa2,Simonucci2015}). The
new results consist in a concrete form of the coefficients of the EFT action,
which are not phenomenological, but they are derived microscopically, starting
from the initial fermionic Hamiltonian. Therefore, when describing the
formation of vortices, the novelty consists in a quantitative description of
the vortex system in a rotating trap.

One of the non-trivial physical results obtained in the present work is the
fact that the vector potential of the rotation in the effective field action
can be different from twice the vector potential for bare fermions. This
difference is due to a renormalization of the effective mass for the pair
field. It is directly related to the fact that the description of an
interacting quantum atomic Fermi gas differs from the known BCS formalism for
superconductors even in the BCS regime, that has been pointed out already in
Ref. \cite{deMelo1993}.

In detail, the rotation leads to a shift in the local chemical potential
$\mu_{\omega}$, and to the appearance of the rotational vector potential
$\mathbf{A}$ in the covariant derivative $-i\nabla_{\mathbf{r}}\rightarrow
-i\nabla_{\mathbf{r}}-\mathring{e}\mathbf{A}$, that leads to the
renormalization factor\ $\tilde{e}$ in the equations of motion for the pair
field. The renormalization factor tends to two in the BEC\ limit, in agreement
with the physical picture of a molecular Bose gas with the boson mass
$m_{p}^{\ast}=2m$. Moving away from the BEC limit, this value diminishes. The
change of the renormalization factor from the BEC limiting value $\tilde{e}=2$
has a clear physical explanation. The fermion pair in a rotating Fermi
condensate moves similarly to a point particle only in the deep BEC regime.
However, beyond the BEC limit, the fermion pair cannot be considered as a
point particle, especially in the BCS regime, where the Cooper-pair size is
large. As a result, the pair effective mass diminishes when the inverse
scattering length moves from BEC to the BCS side.

The renormalization of the effective mass that we obtain is in agreement with
the preceding effective field theory of atomic Fermi gases, as checked by the
comparison of the derived effective field action in particular cases
$T\rightarrow T_{c}$ and $T\rightarrow0$ with reliable works
\cite{deMelo1993,Diener2008,Marini1998,Schakel}. It is also in agreement with
results of the functional renormalization group theory \cite{Diehl}.

Using the obtained formalism, we investigate equilibrium vortex state diagrams
where we identify regions for the superfluid state with no vortices, one
vortex and two or more vortices. For the equilibrium vortex state diagrams in
the variables $\left(  \omega,1/a_{s}\right)  $, the transition curves between
these regions bend over in the BCS regime, in agreement with the results found
using BdG calculations in this regime \cite{Warringa2}. As the number of
particles is increased, the region of the equilibrium vortex state diagram
where vortices are stable extends deeper into the BCS regime. Increasing the
temperature, on the other hand, shrinks the region of stable vortices. The
obtained dependence of the renormalization factor on the inverse scattering
length is essential for these equilibrium vortex state diagrams, especially
for sufficiently weak couplings.

The range of applicability of any kind of the effective field theory
(including, e. g., GL and GP methods) is intrinsically related to the common
assumption for them -- that the order parameter smoothly varies in time and
space. In what concerns the space variation, this means that the EFT can be
applicable when the characteristic scale of the variation of the order
parameter (e. g., the size of vortices or solitons) exceeds the Cooper-pair
correlation length, as discussed in Ref. \cite{Simonucci2014}. The range of
applicability of the present finite temperature EFT has been estimated
quantitatively in Ref. \cite{LAKT2015}. The rotation considered in the present
work does not crucially influence the range of applicability of the EFT.

The equilibrium vortex state diagrams in the variables $\left(  \omega
,T\right)  $ exhibit clear similarity with the results of the BdG method (both
the complete BdG \cite{Warringa,Warringa2} and the coarse graining
approximation for BdG \cite{Simonucci2015}) where a good quantitative
agreement has been found between the critical rotation frequencies obtained
within the present theory and coarse-grained BdG. The lowest critical
frequencies calculated in both EFT and LPDA approaches lie very close to each
other despite the fact that our calculation lies within the NSR-like picture
(where the effective mass of \textquotedblleft dressed\textquotedblright%
\ pairs is renormalized) while the LPDA treatment is in agreement with the
BCS-Leggett picture, where masses of pairs are non-renormalized. This
coincidence is remarkable and may be useful to throw a bridge between these
two paradigms.

We have also arrived at the optimal inverse scattering length for the vortex
formation corresponding to the lowest critical rotation frequency. This value
of the inverse scattering length is in a good agreement with the coupling
strength at which the maximal number of vortices is generated in the
experiment \cite{Zwierlein2005}.

In the present work, we considered the equilibrium configurations of vortices
in rotating traps. The time-dependent phenomena can also be investigated
within the EFT, in general combined with equations for the quasiparticle
distributions. These equations are not an intrinsic part of the EFT and can be
added as an independent ingredient. We have however treated some particular
time-dependent phenomena (travelling dark solitons and collective excitations
in quantum Fermi gases) in Refs. \cite{KTLD2015,KTD2014}, and the KTD
effective field approach appears to be in line with the BdG theory and with experiments.

It is worth noting that an advantage of the present method with respect to the
BdG theory is much shorter computational time and lower memory consumption.
This advantage persists even with respect to the coarse-grained BdG, because
the minimization of the free energy is substantially simpler and faster than a
numerical solution of the differential equations. Moreover, effective field
approaches allow for analytic solutions in many interesting cases, as shown in
our work on dark solitons \cite{KTD2014}. Therefore it is planned to extend
the treatment of non-linear excitations in condensed Fermi gases within the
EFT, involving other factors of interest, such as spin imbalance, two-band
Fermi gases, and spin-orbit coupling. The spin imbalance has been already
incorporated analytically in the coefficients of the effective action
(\ref{SEFT}), and the analysis of effects provided by the imbalance combined
with the rotation is in progress. The spin-orbit coupling will be taken to
account at the microscopic level similarly to Refs. \cite{SO}. Finally, as
shown in Sec. \ref{sec:EFT}, the extension of the present approach to two-band
Fermi gases is straightforward.

The other ingredient which can be incorporated in the EFT is the account of
induced interactions first considered by Gorkov and Melik-Barkhudarov
\cite{GMB}. Their importance for quantum gases in the BCS-BEC crossover was
recently demonstrated \cite{Yu}. The induced interactions lead to substantial
corrections of the parameters of state in the BCS regime, while being less
significant in the BEC regime. Therefore the account of induced interactions
is expected to extend the range of applicability of the EFT towards weak
coupling strengths and to improve a quantitative agreement between EFT and experiment.

\begin{acknowledgments}
We are grateful to G. C. Strinati and H. Warringa for valuable discussions.
This research was supported by the Flemish Research Foundation (FWO-Vl),
project nrs. G.0115.12N, G.0119.12N, G.0122.12N, G.0429.15N, by the Scientific
Research Network of the Research Foundation-Flanders, WO.033.09N, and by the
Research Fund of the University of Antwerp.
\end{acknowledgments}

\appendix

\section{Incorporation of rotation in the effective field theory}

\subsection{Gradient expansion}

The partition function of a fermionic system with two spin states
($\sigma=\uparrow,\downarrow$) is determined by the path integral over the
fermionic fields ,%
\begin{equation}
\mathcal{Z}\propto\int\mathcal{D}\left[  \bar{\psi},\psi\right]  e^{-S}.
\end{equation}
where the action functional $S$ is given by:%
\begin{equation}
S=\int_{0}^{\beta}d\tau\int d\mathbf{r}\left[  \sum_{\sigma=\uparrow
,\downarrow}\bar{\psi}_{\sigma}\left(  \frac{\partial}{\partial\tau}%
+H-\mu_{\sigma}\left(  \mathbf{r}\right)  \right)  \psi_{\sigma}+g\bar{\psi
}_{\uparrow}\bar{\psi}_{\downarrow}\psi_{\downarrow}\psi_{\uparrow}\right]  ,
\end{equation}
where $\beta=1/\left(  k_{B}T\right)  $, $T$ is the temperature, and $k_{B}$
is the Boltzmann constant. To allow for spin imbalance in the Fermi gas,
chemical potentials $\mu_{\sigma}$ are introduced which can be different for
\textquotedblleft spin-up\textquotedblright\ and \textquotedblleft
spin-down\textquotedblright\ species. The coordinate dependent chemical
potentials $\mu_{\sigma}$ are determined by (\ref{mu0}) with $\mu
_{0}\rightarrow\mu_{0,\sigma}$ for each component. The interaction energy with
the coupling constant $g<0$ describes the model contact interactions between
fermions as, for example, in Ref. \cite{deMelo1993}. It represents the Cooper
pairing channel determined by the $s$-wave scattering between two fermions
with antiparallel spins. The one-particle Hamiltonian $H$ in the rotating
frame of reference is determined by formula (\ref{H4}).

After performing the Hubbard-Stratonovich transformation which introduces the
bosonic pair fields $\left(  \bar{\Psi},\Psi\right)  $, and integrating over
the fermionic fields, the partition function becomes \cite{Stoof}%
\begin{equation}
\mathcal{Z}\propto\int\mathcal{D}\left[  \bar{\Psi},\Psi\right]  e^{-S_{eff}},
\label{Z}%
\end{equation}
with the effective bosonic action $S_{eff}$:%
\begin{equation}
S_{eff}=S_{B}-\operatorname{Tr}\left[  \ln\left(  -\mathbb{G}^{-1}\right)
\right]  . \label{Seff1a}%
\end{equation}
We decompose the inverse Nambu matrix $\mathbb{G}^{-1}$ into a sum of the
matrix $\mathbb{F}$ proportional to the pair field $\Psi$, as in Ref.
\cite{KTLD2015},%
\[
\mathbb{F}\left(  \mathbf{r},\tau\right)  =\left(
\begin{array}
[c]{cc}%
0 & -\Psi\left(  \mathbf{r},\tau\right) \\
-\bar{\Psi}\left(  \mathbf{r},\tau\right)  & 0
\end{array}
\right)  ,
\]
and the free-field contribution,
\begin{equation}
\mathbb{G}_{0}^{-1}\left(  \mathbf{r},\tau\right)  =\left(
\begin{array}
[c]{cc}%
-\frac{\partial}{\partial\tau}-H+\mu_{\uparrow} & 0\\
0 & -\frac{\partial}{\partial\tau}+H^{\ast}-\mu_{\downarrow}%
\end{array}
\right)  . \label{Fa}%
\end{equation}
In the momentum representation, the $\mathbb{G}_{0}$ is explicitly obtained
from (\ref{Fa}):%
\begin{equation}
\mathbb{G}_{0}\left(  \mathbf{k},n\right)  =\left(
\begin{array}
[c]{cc}%
\frac{1}{i\omega_{n}-\xi_{\mathbf{k}}+\zeta_{\mathbf{k}}} & 0\\
0 & \frac{1}{i\omega_{n}+\xi_{\mathbf{k}}+\zeta_{\mathbf{k}}}%
\end{array}
\right)  \label{Fa1}%
\end{equation}
with $\xi_{\mathbf{k}}=k^{2}/\left(  2m\right)  -\mu\left(  \mathbf{r}\right)
$ and
\begin{equation}
\zeta_{\mathbf{k}}=\zeta+2\mathbf{k}\cdot\mathbf{A}\left(  \mathbf{r}\right)
,
\end{equation}
where $\mu=\left(  \mu_{\uparrow}+\mu_{\downarrow}\right)  /2$ and
$\zeta=\left(  \mu_{\uparrow}-\mu_{\downarrow}\right)  /2$. As discussed
above, the coordinate-dependent vector potential is taken into account here in
the local density approximation, assuming that $\mathbf{A}\left(
\mathbf{r}\right)  $ varies slowly, as does the trapping potential (which is
included here through the coordinate dependent chemical potential). The above
procedure is quite similar for Fermi gases in three and two dimensions.

Further on, we use the set of units with $\hbar=1$, $2m=1$, the Boltzmann
constant $k_{B}=1$, and the Fermi energy for a free-particle Fermi gas
$E_{F}\equiv\hbar^{2}k_{F}^{2}/\left(  2m\right)  =1$, where $k_{F}%
\equiv\left(  3\pi^{2}n\right)  ^{1/3}$ is the Fermi wave vector and $n$ is
the fermion density. Therefore in the present work, $k_{F}=1$, and the lengths
are measured in units of $1/k_{F}$.

The next step is the gradient expansion of the effective action (\ref{Seff1a})
following exactly the same scheme as in Ref. \cite{KTLD2015}, up to the
second-order derivatives in time and in space. A complete summation in powers
of the squared amplitude of the pair field $w\equiv\left\vert \Psi\right\vert
^{2}$ is performed in each term of this gradient expansion separately. As a
result, the following effective field action is obtained, which is
structurally similar to that derived in Ref. \cite{KTLD2015} but with a new
term provided by rotation:%
\begin{align}
S_{eff}  &  =\int_{0}^{\beta}d\tau\int d\mathbf{r}\left\{  \left[  \Omega
_{s}\left(  w\right)  +\frac{D}{2}\left(  \bar{\Psi}\frac{\partial\Psi
}{\partial\tau}-\frac{\partial\bar{\Psi}}{\partial\tau}\Psi\right)  \right.
\right. \nonumber\\
&  +Q\frac{\partial\bar{\Psi}}{\partial\tau}\frac{\partial\Psi}{\partial\tau
}-\frac{R}{2w}\left(  \frac{\partial w}{\partial\tau}\right)  ^{2}+C\left(
\nabla_{\mathbf{r}}\bar{\Psi}\cdot\nabla_{\mathbf{r}}\Psi\right)  -E\left(
\nabla_{\mathbf{r}}w\right)  ^{2}\nonumber\\
&  \left.  \left.  +iG\mathbf{A}\cdot\left(  \bar{\Psi}\nabla_{\mathbf{r}}%
\Psi-\Psi\nabla_{\mathbf{r}}\bar{\Psi}\right)  \right]  \right\}  . \label{S3}%
\end{align}
The coefficients in this effective field action (generalized here for a $\nu
$-dimensional Fermi gas with $\nu=2,3$) take the form:%
\begin{align}
C  &  =2\int\frac{d^{\nu}\mathbf{k}}{\left(  2\pi\right)  ^{\nu}}\frac{k^{2}%
}{\nu}f_{2}\left(  \beta,E_{\mathbf{k}},\zeta_{\mathbf{k}}\right)
,\label{c}\\
D  &  =\int\frac{d^{\nu}\mathbf{k}}{\left(  2\pi\right)  ^{\nu}}\frac
{\xi_{\mathbf{k}}}{w}\left[  f_{1}\left(  \beta,\xi_{\mathbf{k}}%
,\zeta_{\mathbf{k}}\right)  -f_{1}\left(  \beta,E_{\mathbf{k}},\zeta
_{\mathbf{k}}\right)  \right]  ,\label{d}\\
E  &  =\frac{4}{\nu}\int\frac{d^{\nu}\mathbf{k}}{\left(  2\pi\right)  ^{\nu}%
}k^{2}~\xi_{\mathbf{k}}^{2}~f_{4}\left(  \beta,E_{\mathbf{k}},\zeta
_{\mathbf{k}}\right)  ,\label{ee}\\
Q  &  =\frac{1}{2w}\int\frac{d^{\nu}\mathbf{k}}{\left(  2\pi\right)  ^{\nu}%
}\left[  f_{1}\left(  \beta,E_{\mathbf{k}},\zeta_{\mathbf{k}}\right)  \right.
\nonumber\\
&  \left.  -\left(  E_{\mathbf{k}}^{2}+\xi_{\mathbf{k}}^{2}\right)
f_{2}\left(  \beta,E_{\mathbf{k}},\zeta_{\mathbf{k}}\right)  \right]
,\label{qq}\\
R  &  =\int\frac{d^{\nu}\mathbf{k}}{\left(  2\pi\right)  ^{\nu}}\left[
\frac{f_{1}\left(  \beta,E_{\mathbf{k}},\zeta_{\mathbf{k}}\right)  +\left(
E_{\mathbf{k}}^{2}-3\xi_{\mathbf{k}}^{2}\right)  f_{2}\left(  \beta
,E_{\mathbf{k}},\zeta_{\mathbf{k}}\right)  }{3w}\right. \nonumber\\
&  \left.  +\frac{4\left(  \xi_{\mathbf{k}}^{2}-2E_{\mathbf{k}}^{2}\right)
}{3}f_{3}\left(  \beta,E_{\mathbf{k}},\zeta_{\mathbf{k}}\right)
+2E_{\mathbf{k}}^{2}wf_{4}\left(  \beta,E_{\mathbf{k}},\zeta_{\mathbf{k}%
}\right)  \right]  . \label{rr}%
\end{align}
The functions $f_{p}\left(  \beta,\varepsilon,\zeta\right)  $ have been
introduced in Ref. \cite{KTLD2015}. They are defined through fermionic
Matsubara sums,%
\begin{equation}
f_{p}\left(  \beta,\varepsilon,\zeta\right)  \equiv\frac{1}{\beta}%
\sum_{n=-\infty}^{\infty}\frac{1}{\left[  \left(  \omega_{n}+i\zeta\right)
^{2}+\varepsilon^{2}\right]  ^{p}}, \label{fp}%
\end{equation}
and have been expressed explicitly using the recurrence relations:%
\begin{align}
f_{1}\left(  \beta,\varepsilon,\zeta\right)   &  =\frac{1}{2\varepsilon}%
\frac{\sinh(\beta\varepsilon)}{\cosh(\beta\varepsilon)+\cosh(\beta\zeta
)},\label{msum}\\
f_{p+1}\left(  \beta,\varepsilon,\zeta\right)   &  =-\frac{1}{2p\varepsilon
}\frac{\partial f_{p}\left(  \beta,\varepsilon,\zeta\right)  }{\partial
\varepsilon}.
\end{align}
The coordinate-dependent thermodynamic potential for a rotating Fermi gas is
determined by the expressions:%
\begin{align}
\Omega_{s}\left(  w\right)   &  =-\int\frac{d\mathbf{k}}{\left(  2\pi\right)
^{3}}\left(  \frac{1}{\beta}\ln\left(  2\cosh\beta E_{\mathbf{k}}+2\cosh
\beta\zeta_{\mathbf{k}}\right)  \right. \nonumber\\
&  \left.  -\xi_{\mathbf{k}}-\frac{w}{2k^{2}}\right)  -\frac{w}{8\pi a_{s}%
}\quad\left(  \text{in 3D}\right)  , \label{Ws}%
\end{align}
and%
\begin{align}
\Omega_{s}\left(  w\right)   &  =-\int\frac{d^{2}\mathbf{k}}{\left(
2\pi\right)  ^{2}}\left(  \frac{1}{\beta}\ln\left(  2\cosh\beta E_{\mathbf{k}%
}+2\cosh\beta\zeta_{\mathbf{k}}\right)  \right. \nonumber\\
&  \left.  -\xi_{\mathbf{k}}-\frac{w}{2k^{2}+E_{b}}\right)  \quad\left(
\text{in 2D}\right)  , \label{Ws1}%
\end{align}
where $E_{b}$ is the binding energy for a two-particle bound state in 2D.

Finally, when performing the gradient expansion, rotation leads to a new term
in the effective field action (\ref{S3}), proportional to the first-order
space gradient of the pair field,%
\begin{equation}
\delta S_{eff}^{\left(  rot\right)  }=\int_{0}^{\beta}d\tau\int d\mathbf{r}%
~iG\mathbf{A}\cdot\left(  \bar{\Psi}\nabla_{\mathbf{r}}\Psi-\Psi
\nabla_{\mathbf{r}}\bar{\Psi}\right)  . \label{dSrot}%
\end{equation}
In the absence of rotation, this term vanishes due to inversion symmetry. It
is calculated as in Ref. \cite{KTLD2015}, summing up the whole series in
powers of the amplitude of the pair field in the coefficients at $\nabla\Psi$
and $\nabla\bar{\Psi}$. The new coefficient $G$, which appears due to the
rotation, is:%
\begin{align}
G  &  =D\nonumber\\
&  +\int\frac{d^{\nu}\mathbf{k}}{\left(  2\pi\right)  ^{\nu}}\frac{1}{w}%
\frac{\left(  \mathbf{k\cdot A}\right)  }{\left\vert \mathbf{A}\right\vert
^{2}}\zeta_{\mathbf{k}}\left[  f_{1}\left(  \beta,\zeta_{\mathbf{k}}%
,\xi_{\mathbf{k}}\right)  -f_{1}\left(  \beta,\zeta_{\mathbf{k}}%
,E_{\mathbf{k}}\right)  \right]  . \label{G}%
\end{align}

In summary, the effect of rotation on the effective field action functional
derived in Ref. \cite{KTLD2015} is taken into account through the
renormalization of the averaged chemical potential $\mu$ according to
(\ref{mu0}) and the replacement of the chemical potential imbalance as
$\zeta\rightarrow\zeta_{\mathbf{k}}$. This may create a wrong impression that
rotation can lead to polarized Fermi gases at $\zeta=0$. However, this is not
the case. For clarity, let us consider a comparison between the real
electromagnetic vector potential and the rotational vector potential. A real
electromagnetic vector potential for particles with a true spin will lead to
Zeeman splitting for spin states, so the chemical potentials of the two
components can be different. The Zeeman splitting of \textquotedblleft
spin\textquotedblright\ states for atomic Fermi gases due to rotation is, in
general, absent. On the contrary, splitting for the momentum states due to
rotation occurs in the same way as due to a magnetic field \cite{Alben1969}.
Moreover, this local-momentum splitting of the chemical potential appears in
the Nambu tensor in the same way as in the Nambu-Gorkov theory. In order to
see this, we can refer to the works \cite{Warringa,Warringa2,Urban2008}, where
the inverse Nambu matrix appears with the same one-particle Hamiltonian as in
the present work. However, for a balanced gas, the contributions with
$\zeta_{\mathbf{k}}$ and $\zeta_{-\mathbf{k}}$ cancel out in the integration
over $\mathbf{k}$, and hence rotation does not lead to a population imbalance.

The appearance of the local-momentum splitting of the chemical potential\ is
physically transparent. In a Cooper pair, the two fermions have opposite
momenta. In the presence of rotation, their single-particle energies become
unequal, in the same way as two pairing electrons in the magnetic field
experience a Zeeman splitting. Note that for Cooper-paired electrons in a
magnetic field, the Lorentz force destabilizes the pair already at much lower
magnetic field than that where the Zeeman splitting breaks up the pair --
however, for the neutral atoms, this effect is absent.

This physical picture assumes that the Cooper pair size is small with respect
to a characteristic size of the superfluid system (for example, the radius of
the trap) so that the background parameters within the extent of a Cooper pair
are approximately uniform. This condition needs to be fulfilled in order for
any description in terms of an effective field theory
\cite{Nishida,Marini1998,Schakel,KTD2014,Simonucci2014} to be applicable. It
should be noted that whereas the aforesaid splitting of the fermion energy is
a standard result for the Bogoliubov - de Gennes theory, it has not been taken
into account in existing effective field theories, so that this seems to be
new with respect to other EFT-like approaches.

In accordance with Ref. \cite{Schakel}, $G=D$ in (\ref{G}) corresponds to the
leading order and the term in the second line corresponds to the
next-to-leading order in the effective field theory. Also the splitting
$\zeta_{\mathbf{k}}$ of the chemical potential is the next-to-leading order
correction with respect to the renormalization of $\mu$ due to rotation. Hence
these corrections must be relatively small \emph{within the range of
applicability of the effective field theory}. Moreover, they should be
neglected for consistency, because they may lead to non-controlled corrections
beyond EFT.

A question may appear whether next-to-leading order terms can be important
near a vortex core, where the order parameter rises rapidly. The range of
applicability of the leading-order approximation is in fact the same as the
range of applicability of any other effective field theory, e. g., the
Ginzburg-Landau equation which is often used for the analysis of the vortices
in superconductors and superfluids. This question is more general than the
subject of the present study, because it is the same for rotating and
non-rotating superfluids. It was studied in Refs. \cite{KTLD2015,LAKT2015} by
a comparison of the obtained vortex parameters with results of the alternative
microscopic approach -- the BdG theory.

We can also show that next-to-leading order corrections should be neglected in
order to satisfy the gauge invariance for the effective field action. In the
derivation above, we start from the action for the fermionic field $\psi$ in
the lab frame, then transform it to the rotating frame of reference, and
finally perform the Hubbard-Stratonovich transformation to introduce the pair
field $\Psi$. As a check of the gauge invariance of the obtained effective
field action, we also consider inverting the order of these operations: first
obtaining the action for the pair field $\Psi$ in gradient expansion, and then
applying the transformation to the rotating frame of reference. In that case,
the energy term $-\omega L_{z}$ (where $\hat{L}_{z}$ is the $z$ component of
the orbital angular momentum for the pair field $\Psi$) appears in the bosonic
pair Hamiltonian directly from the condition of the gauge invariance --
similarly as in the Gross-Pitaevskii theory \cite{Stoof}. This order of
operations leads to the same final result as obtained above (\ref{SEFT}%
)-(\ref{CD}), but with the coefficient $G=D$. The resulting effective field
action takes then the form:%
\begin{align}
S_{eff}  &  =\int_{0}^{\beta}d\tau\int d\mathbf{r}\left\{  \left[  \Omega
_{s}\left(  w\right)  +\frac{D}{2}\left(  \bar{\Psi}\frac{\partial\Psi
}{\partial\tau}-\frac{\partial\bar{\Psi}}{\partial\tau}\Psi\right)  \right.
\right. \nonumber\\
&  +Q\frac{\partial\bar{\Psi}}{\partial\tau}\frac{\partial\Psi}{\partial\tau
}-\frac{R}{2w}\left(  \frac{\partial w}{\partial\tau}\right)  ^{2}+C\left(
\nabla_{\mathbf{r}}\bar{\Psi}\cdot\nabla_{\mathbf{r}}\Psi\right)  -E\left(
\nabla_{\mathbf{r}}w\right)  ^{2}\nonumber\\
&  \left.  \left.  +iD\mathbf{A}\cdot\left(  \bar{\Psi}\nabla_{\mathbf{r}}%
\Psi-\Psi\nabla_{\mathbf{r}}\bar{\Psi}\right)  \right]  \right\}  \label{S4}%
\end{align}
The coefficients $D,C,E,Q,R$ in this effective field action are the same as in
Ref. \cite{KTLD2015}. The new term $\left(  \propto\mathbf{A}\right)  $
expresses the coupling of the rotational vector potential to the current density.

\subsection{Renormalization of the pair mass}

The terms with the gradient of the pair field can be equivalently rewritten in
terms of the covariant derivatives,
\begin{align}
&  \int d\mathbf{r}\left[  C\left(  \nabla_{\mathbf{r}}\bar{\Psi}\cdot
\nabla_{\mathbf{r}}\Psi\right)  +iD\mathbf{A}\cdot\left(  \bar{\Psi}%
\nabla_{\mathbf{r}}\Psi-\Psi\nabla_{\mathbf{r}}\bar{\Psi}\right)  \right]
\nonumber\\
&  =\int d\mathbf{r}\left[  C\left\vert \left(  \nabla_{\mathbf{r}}%
-i\mathring{e}\mathbf{A}\right)  \Psi\right\vert ^{2}-C\mathring{e}^{2}%
A^{2}\left\vert \Psi\right\vert ^{2}\right]  , \label{CD}%
\end{align}
with the renormalization factor $\mathring{e}=D/C$. As established in Ref.
\cite{KTLD2015} the coefficient $D$ enters the equations of motion for the
pair fields only through the combination $\tilde{D}\equiv\partial\left(
wD\right)  /\partial w$. Consequently, physical sense can be attributed to the
other renormalization factor,
\begin{equation}
{\tilde{e}}=\frac{1}{C}\frac{\partial\left(  wD\right)  }{\partial w}.
\label{etild1}%
\end{equation}

The physical sense of the renormalization factor $\tilde{e}$ can be explained
using the following reasoning. Let us temporarily, just for illustration
purposes, neglect the terms with coefficients $E,Q,R$ (which are not necessary
for this explanation) in the EFT action. In the absence of rotation, the
equation of motion for the pair field in the real-time representation
(simplifying the equation of motion from Ref. \cite{KTLD2015}) then becomes:%
\begin{equation}
i\frac{\partial\Psi}{\partial t}=-\frac{1}{2m_{p}^{\ast}}\nabla_{\mathbf{r}%
}^{2}\Psi+\frac{1}{\tilde{D}}\frac{\partial\Omega_{s}}{\partial w}\Psi,
\label{eqmot}%
\end{equation}
with the effective mass of the pair%
\begin{equation}
m_{p}^{\ast}\equiv\frac{\tilde{D}}{2C}.
\end{equation}
This equation is similar to the Gross-Pitaevskii one, and is exactly reduced
to the GP form if we expand the thermodynamic potential in powers of
$\left\vert \Psi\right\vert ^{2}$ up to the second order. In general,
$m_{p}^{\ast}\neq1$. This result is not surprising, because a renormalization
of the effective pair mass with respect to twice the fermion mass can be
straightforwardly obtained from the effective field actions of earlier works,
e. g., Refs. \cite{deMelo1993,Huang}. Note that in Ref. \cite{Huang} it is
explicitly stated that the effective boson mass is equal to unity only in the
BEC limit. Moreover, the renormalization of the coefficients at the space
gradients and time derivatives is predicted by the EFT formulated using the
functional renormalization group method \cite{Boettcher,Diehl}.

The rotation can be incorporated in the GP-like equation (\ref{eqmot}) in the
same way as in the Schr\"{o}dinger equation -- considering the Bose gas of
pairs which is at rest in the rotating frame of reference. In the same way as
described above for fermions, the rotation applied to (\ref{eqmot}) leads to
the appearance of the rotational\ vector potential for the pair field%
\begin{equation}
\mathbf{A}_{p}\left(  \mathbf{r}\right)  =m_{p}^{\ast}\left[
\boldsymbol{\omega}\times\mathbf{r}\right]  .
\end{equation}
Thus the renormalization factor $\tilde{e}=2m_{p}^{\ast}$\ has the physical
sense of the renormalized effective mass for the pair field in units of the
fermion mass.

\end{document}